\documentclass[prl,twocolumn,floatfix,showpacs,showkeys]{revtex4} 

\usepackage[first,draft]{draftcopy}
\usepackage{latexsym,amssymb,amsmath,amsfonts}
\usepackage{graphicx}
\usepackage{bm}
\newcommand{\beq}{\begin{equation}}
\newcommand{\eeq}{\end{equation}}
\newcommand{\bc}{\begin{center}}
\newcommand{\ec}{\end{center}}
\newcommand{\eeqa}{\end{eqnarray}}
\newcommand{\beqa}{\begin{eqnarray}}

\newcommand{\ra}{\rightarrow}

\newcommand{\al}{\alpha}
\newcommand{\be}{\beta}
\newcommand{\ga}{\gamma}

\newcommand{\de}{\delta}

\newcommand{\ep}{\epsilon}

\newcommand{\la}{\lambda}

\newcommand{\si}{\sigma}

\newcommand{\ta}{\tau}

\newcommand{\ph}{\phi}

\newcommand{\ed}{\end{document} }

\begin{document}

\title{Stringy Maxwell charge and the magnetic dipole moment}
\author{Richard T. Hammond}

\email{rhammond@email.unc.edu }
\affiliation{Department of Physics\\
University of North Carolina at Chapel Hill\\
Chapel Hill, North Carolina and\\
Army Research Office\\
Research Triangle Park, North Carolina}

\date{\today}


\begin{abstract}
It is shown that stringy charge leads naturally to an observable magnetic moment of the electron.
\end{abstract}

\maketitle
   
\section{Introduction}

The point model of the electron has been very successful in describing electromagnetic effects. However, over the years there have been a number of attempts to build a theory based upon an extended model. For example, Uhlenbeck and Goudsmit envisioned the electron as a spinning shell of charge to account for the magnetic dipole moment of the electron (which was necessary to explain the Zeeman shifts). However, Lorentz noted the self energy would be wrong, in other words, the shell cannot spin fast enough.\cite{uhlenbeck} Soon after, however, the Dirac equation put electron spin on a firm theoretical foundation, and the notion of a classical model of the magnetic moment faded.

Other reasons for a non-zero sized electron arose from the problem of radiation reaction. The Lorentz-Abraham-Dirac equation predicts solutions in which a free electron at rest suffers an extreme acceleration, sending it to nearly the speed of light in less than a zeptosecond. It was found if the point model of the electron is replaced by a finite size electron, then the runaway solutions could be avoided. A review of these effects may be found in the literature.\cite{hammondrr}

An extended model of the electron has been used in general relativistic models as the source of the Kerr-Newman field.\cite{lopez} Burinskii also argues in favor of an extended model in terms of strings.\cite{burinskii}

In general, string theory allows for structure. In this case particles are described by one-dimensional objects, and so, one may wonder if such a structure may account for the magnetic dipole moment. At first glance one may be skeptical. After all, the spinning shell of Uhlenbeck and Goudsmit did not work, and a string is much, much smaller than their shell. But, as we will see, there is more to the story and, in fact, it will be shown a string model of the electron can indeed account for its magnetic moment.

For the reasons stated above it is often stated that classical physics cannot account for the magnetic moment of the electron. This is not true. What is true is that a classical spinning shell cannot account for the magnetic moment. Fortunately, in the quantum mechanical realm, the canonical momentum $p^\mu-eA^\mu$ in the Dirac equation gives the correct magnetic interaction, so the problem just described in sidestepped.

Closely related to the magnetic moment is the intrinsic spin of a particle. These quantities are not inseparable however as the neutrino shows. It was long believed intrinsic spin could not be accounted for classically but this changed when it was realized general relativity, taken either  a local gauge theory or of the string theory type, describes classical intrinsic spin.\cite{hehl}\cite{hammondtg} In particular it was shown a closed string gives rise to  spin due to its structure, and not because of its motion.\cite{shs} It was proven this was indeed intrinsic spin from an analysis of the equation of motion, which follows from the Bianchi identities.

If a string can account for the observed spin of a particle, then it is reasonable to ask if a charged string can give rise to the magnetic moment.
Note the charge $e$ of an electron is  put into the classical action and is indeed described by classical physics. The magnetic moment is equally fundamental so it is reasonable to suspect classical physics can describe it.

It is not simply an academic question. Laser intensities of up to 10$^{22}$W cm$^{-2}$ have been achieved,\cite{yavonsky} and the spin effects should be included in the equations of motion.\cite{walser} These are highly classical effects, and in order to obtain the correct equations, one must start with the correct model of the electron.

To be concrete, let us consider the action principle from which electrodynamics follows. It is  $\de I=0$ where

\beq\label{emaction}
I=-m\int ds-\frac{1}{16\pi}\int d^4x F_{\mu\nu}F^{\mu\nu}
+e\int A_\mu dx^\mu
\eeq
where $dx^\mu$ runs along the worldline of the particle parameterized by $s$. The third term on the right necessarily makes the electron a point-like object

Besides the issue of the equation of motion of particles with charge and a magnetic dipole moment, there is a logical issue at stake.
Electromagnetism is often touted as one of the best theories we have. It is renormalizable, and QED measurements of $g-2$ have shown agreement with theory to one part in a trillion.\cite{odom} There is one situation, however, in which this formulation is inadaquate. To see this, consider the field of a electron at the origin in its rest frame. There is the electric field, which arises from the zero component of the current density $j^\mu$, which is $j^0=e\de(x)$. There is also a magnetic field due to the magnetic dipole moment of the electron, but this cannot be accounted for by $j^\mu$, as is well known, even though we know the moment exists.

Consider the example a spin polarized material (with zero orbital angular moment) giving rise to a macroscopic magnetic field that can be measured classically. This field is simply the vector sum of the intrinsic dipoles. This yields a measurable field, yet (\ref{emaction}) cannot account for the field. This is because the action (\ref{emaction}) insures the charge is a point-like object and in fact, is given in terms of a delta function.

If we consider elementary particles to be strings, then it is natural to consider charge as being spread along the string (stringy charge). 
As an example, consider mass. The matter term is generalized by

\beq
m\int ds \ra \mu\int \sqrt{-\ga}d^2\zeta
\eeq
where $\ga_{ab}$ is the induced metric in string space and $\ga$ its determinant. We may interpret $\mu$ as the mass per unit length, so this shows a natural generalization from a point mass to a stringy mass. However, in the electromagnetic case, there is no natural coupling that accomplishes this and so, in conventional string theory, the charge is anchored at the end of the string as a point which moves along the brane.

However, strings endowed with charge per unit length in four dimensions have been under investigation for some time. Strings in anti-de Sitter spacetime have been studied in the case that the charge arises from rotation.\cite{dias} Related result were found in Gauss-Bonnet gravity with a nonlinear Maxwell source,\cite{hendi} and charge on a hyperplane was found in Chern-Simons gravity.\cite{bunster} Charged strings in higher  dimensions have also ben considered,\cite{lee} as well as smeared charge effects.\cite{armas}

String charge has been considered by Vilenkin  and Shellard by introducing a scalar field $\phi(\zeta^0,\zeta^1)$ where ${\zeta^0,\zeta^1}$ are the string coordinates.\cite{vilenkin} In this case the action is

\begin{widetext}
\beq\label{action}
I=-\mu\int \sqrt{-\ga}d^2\zeta-\frac{1}{16\pi}\int d^4x\sqrt{-g}
F_{\mu\nu}F^{\mu\nu}
+\int \sqrt{-\ga}d^2\zeta( A_\mu\ph,_\nu X^\mu,_a X^\nu,_b\ep^{ab}-\frac12\ph,_a\ph,_b\ga^{ab}
)
.\eeq
\end{widetext}
The Latin letters $a, b$ run from 0 to 1, $\sqrt{-\ga}\ep^{ab}$ is the 2-D antisymmetric tensor, and $X^\mu=x^\mu(\zeta^0,\zeta^1)$. The third term generalizes the third term of (\ref{emaction}) and the final term is the kinetic term for the scalar field. The electric charge $e$ is present in the third term because, as we shall see, it is embedded in $\phi$.
The question is, can we expect a stringy formulation to endow the string with a magnetic dipole moment within the classical realm.

We will work in the conformal gauge and set $\zeta^0 = t$.
To begin, variations of (\ref{action}) are taken with respect to $x^\mu$, $A_\mu$, and $\ph$, which give

\beq
F^{\nu\mu}_{\ \ ;\nu}=4\pi j^\mu
\eeq
where

\beq\label{j}
j^\mu=\frac{1}{\sqrt{-g}}\int d^2\zeta\sqrt{-\ga}\de^4\ph,_b x^{\mu}_{ ,a}\ep^{ab}
,\eeq

\beq\label{boxx}
\mu\Box X^\nu=
\frac{1}{\sqrt{-\ga}}(\sqrt{-\ga}X^\nu,_a t^{ab}),_b
+F^\nu_{\ \mu}j^\mu
,\eeq
where

\beq
t^{ab}=\ph^{,a}\ph^{,b}-\frac12 \ga^{ab}\ph,_n\ph^{,n}
\eeq

and

\beq\label{boxph}
\Box\ph=\frac 1 2F_{\mu\nu} X^\mu_{ ,a}X^\nu_{ ,b}\ep^{ab}
.\eeq

Now consider the case the string coupling dominates and we may set the right side of (\ref{boxph}) and (\ref{boxx}) to zero.\cite{vilenkin} 
A solution to (\ref{boxx}) (with the right side set to zero) for a closed string is

\beq
X^\mu=a\ta+i\sqrt{\frac{\al'}2}\sum_{n\ne0}n^{-1}e^{-in\ta}(\al_n e^{in\si}+\tilde\al_n e^{-in\si})
\eeq
where the $\al_n$ and $\tilde\al_n$ are subject to the Virasoro constraints. A simple solution in which the string is contained in the $x-y$ plane is given by\cite{kibble}

\beq
{\bm X}= \frac{L}{4\pi}\left(
(\sin{\chi_{-}}+\sin{\chi_{+}})\hat{\bm x}
-(\cos{\chi_{-}}+\cos{\chi_{+}})\hat{\bm y}
\right)
\eeq
where $\chi_{\pm}\equiv 2\pi(\zeta \pm t)/L$.

The string is circular but time dependent. It starts from its maximum radius and then shrinks to a point, at which time the current changes direction, and expands to its maximum. On average, the current is zero. However magnetic moments, which are integrals of distance times the current, though time dependent, need not average to zero.

The solution to $\Box_2\phi=0$ may be written in spectral form,
\beq\label{phi}
\phi=  \al t+ \be \zeta+
 \sum_n\left(
a_n\sin{(n\chi_{+}+\la_n)} +b_n\cos{(n\chi_{-}+\la_n)}
\right)
\eeq
where $\la_n$ is a phase constant.
Using (\ref{j}) we have an expression for the current density of the string, which is

\beq
j^\mu=\int d^2\zeta\de^4(\ph,_1X^\mu,_0-\ph,_0X^\mu,_1)
.\eeq
For the zero component, Since $X^0,_1=0$, we obtain
\beq
j^0=e\de^3
\eeq
where we used $\be\int d\zeta=\be L\equiv e$ where $L$ is the length of the string. This has the interpretation $\be$ equals the charge per unit length along the string (the electric dipole moment of this string is zero). Note that in this formulation $\phi$ has the dimension of electric charge, and the action has the correct dimnesions of $e^2$ (or $\hbar$ with $c=1$).

More interesting is the  expression for $j^n$, the spatial part of the current density, 

\beq\label{jn}
j^n=\int d^2\zeta\de^4(\ph,_1X^n,_0-\ph,_0X^n,_1)
.\eeq
The potential may be found from,

\beq
A^\mu(x)=\int d^4x' G(x-x')j^\mu(x')
\eeq
where $G(x-x')$ is the retarded Green's function, so that

 \beq
A^\mu(x)=\int d^4x' \frac{j^\mu(x')}{|\bf x - \bm X|}\de(t'-|\bf x - \bm X|)
.\eeq
With (\ref{j}) this becomes

 \beq
A^\mu(x)=\int d^4x' d\zeta\frac{\de^3(\bf x' - \bm X)
\ep^{ab}\ph,_b x^\mu,_a\de(t'-t-|\bf x' - \bm X|)}
{|\bm x - \bm x'|}
.\eeq

We may expand in the usual way if $x>>x'$,
and performing the integral over $dt'$ we have

 \beq
A^\mu(x)=\int d^3x' d\zeta\frac{\de^3N^\mu
}
{\psi}
\left( \frac1x+\frac{\bm x \cdot \bm x'}{x^3}\right)
\eeq
where $\psi=1+\hat {\bm R}\cdot { \dot{\bm x'}}$ and
$N^\mu=\ep^{ab}\ph,_b x^\mu,_a$.

Not let us consider the average over the retarded time $\ta$. Using
$\ta=r-|\bm x-\bm x'|$, and integrating over the delta function we have

\beq
\bm A=\frac{\bm\mu\times \bm x}{x^3}
\eeq
where

\beq
\bm\mu=\frac1T\int \bm f\times\bm N d\zeta dt
\eeq
where now it is understood we are taking the average value, and it is noted the first term in the expansion vanishes (note that $T=L$). Integrating over time and then over $\zeta $
finally yields

\beq
\bm\mu=-\frac{\al L^2}{8\pi} \hat{\bm z}
\eeq
which shows $\al$ is of dimension magnetic dipole per unit area.

This shows stringy charge gives rise to a magnetic dipole moment. Although the string is not static, the constant magnetic  dipole moment does not arise from any sort of charge times velocity. In fact it is independent of the charge as the formulation shows.
Viewing (\ref{phi}), we see the charge is proportional to $\beta$ and the magnetic moment is proportional to $\al$, two independent constants. However, this model does not predict the value of the magnetic dipole moment, nor does it predict the charge. These quantities are still found by experiment.

\ed

\item R. T. Hammond, ``Spin, the classical theory,'' J. Mod. Phys., 1 {\bf 3} (2012).

Below we are interested in the time average results so that, at a fixed distant point we may write

\beq A_n(x)=\frac jx \int j_n(x')d^3x'+\frac{\bm x}{ x^3}\cdot \int {\bm x'}j_n(x')d^3x'+...
,\eeq
where the time is the retarded value and since $j_n$ is localized, we write the magnetic dipole part as

\beq\label{A}
{\bm A}
	=\frac{{\bm \mu}\times{\bm r}}{ r^3}
	\eeq
where

\beq
{\bm \mu}=
\frac12\int{\bf r}(x')\times{\bf j}(x')dV'
\eeq
which becomes ${\bm \mu}=\mu\hat{\bm z}$

\begin{widetext}
\beq
\mu=\frac{L}{4\pi}\int d\zeta dx'dy'\de(x'-x(\zeta))\de(y'-y(\zeta))
\Big[
\left(\sin{\chi_{-}}+\sin{\chi_{+}}
\right)
(\ph,_1X^2,_0-\ph,_0X^2,_1)-
(\cos{\chi_{-}}+\cos{\chi_{+}})
(\ph,_1X^1,_0-\ph,_0X^1,_1)
\Big]
.\eeq
\end{widetext}

Integrating over the prime variables first, and then over $\zeta$ we find